# The Puzzle of Meteoritic Minerals Heideite and Brezinaite; Are they Iron-based Superconductors? Are they Technosignatures?

*B. P. Embaid*

*Laboratorio de Magnetismo, Escuela de Física, Universidad Central de Venezuela, Apartado 47586, Los Chaguaramos, Caracas 1041-A, Venezuela.*

## Abstract

Transition metal sulfides $(Fe, V)_3S_4$ and $(Fe, Ti)_3S_4$, with the monoclinic $Cr_3S_4$ type structure have been studied for a long time, itinerant magnetism in form of Spin density waves (SDW) have been found in these systems with different features as evidenced by $^{57}Fe$ Mössbauer Spectroscopy, there is an intricate relationship between the proportion of Fe, V and Ti atoms, the degree of commensurability of the SDW and the magnetic transition temperature. These sulfides have no natural occurrence on Earth and some of these phases were detected as minerals in meteorites; the mineral Heideite in the Bustee and Kaidun meteorites, with minor proportion of Cr atoms leading to the general formula $(Fe, Cr)_{1+x} (Ti, Fe)_2 S_4$, and the mineral Brezinaite in the Tucson meteorite, with minor proportion of Fe atoms and traces of V, Ti and Mn atoms, leading to the formula $(Cr_{2.65}Fe_{0.20}V_{0.09}Ti_{0.06}Mn_{0.04})_{3.04}S_4$.

In this critical review of the experimental literature, we discuss the issues that these meteoritic minerals are structurally sensitive to the method of synthesis, and so is their magnetic behavior, especially in the presence of minor and trace elements. This discussion could shed light on our knowledge in Solid State Physics and Planetary Science; these meteoritic minerals are promising candidates for iron-based superconductors because of three clues: they are layered structures, they undergo a transition to SDW with variable degree of commensurability and the minor and trace elements could act as dopants and hence suppress the SDW giving rise to superconductivity. On the other side, the genesis of these meteoritic minerals could require controlled and sophisticated process not easily found in nature. So, it is important to be open-minded and even provocative to consider the following question: Are these meteoritic minerals samples of Extraterrestrial Technosignatures?

## 1. INTRODUCTION

In this manuscript we present a summary of experimental results on Solid State Physics properties of synthetic transition metal sulfides, and discuss their implication in Planetary Science related to the mineralogical properties that appear to be present in meteoritic minerals with similar structure and slight variation in composition.

More specifically, we will talk about the systems (Fe, V)$_3$S$_4$ and (Fe, Ti)$_3$S$_4$; their structural, magnetic and transport properties, and two meteoritic minerals; the mineral Heideite (Fe, Cr)$_{1+x}$ (Ti, Fe)$_2$S$_4$ with occurrence in the meteorites Bustee (India) and Kaidun (Yemen) and the mineral Brezinaite (Cr$_{2.65}$Fe$_{0.20}$V$_{0.09}$Ti$_{0.06}$Mn$_{0.04}$)$_{3.04}$S$_4$ with occurrence in the Tucson meteorite (USA).

Based on experimental results reported so far for synthetic sulfide systems, we will identify open issues in these meteoritic minerals. These issues deserve future experiments in the attempt to be solved; they are categorized in two types; a) Solid State Physics issue, focused on the structural, magnetic and transport behaviors and b) Planetary Science issue, focused on the genesis.

The significance to solve these issues relies in their impact on our knowledge in these two disciplines, that is, in Solid State Physics these minerals are promising candidates for iron-based superconductors, and in Planetary Science the synthesis method leading to analog samples of these minerals could require controlled and sophisticated process not easily found in nature. Although it is of course possible that there are as-yet-understood processes that can lead to the formation of these minerals, however, we can step ahead on the possible new interpretation in the recently emerging field: Technosignatures.

The redaction of the manuscript is adapted to be comprehensive from the experimental point of view, and it can be readable by audience from other disciplines of interest like Planetary Science and Astrobiology. There is a minimum of technical terms used in Solid State characterization techniques involved, such as $^{57}$Fe Mössbauer spectroscopy and Crystallography, which are necessary for peer review purpose.

The manuscript will be divided into the following sections, briefly described below:

2. Selected transition metal sulfides.
3. Meteoritic minerals Heideite and Brezinaite.
4. Why are these minerals so interesting?
5. Significance in Solid State Physics and Planetary Science.
6. Conclusions.

Sections 2 and 3 are focused on our current knowledge in Solid State characterization of synthetic systems (Fe, V)$_3$S$_4$ and (Fe, Ti)$_3$S$_4$ and the meteoritic minerals Heideite (Fe, Cr)$_{1+x}$ (Ti, Fe)$_2$S$_4$ and Brezinaite (Cr$_{2.65}$Fe$_{0.20}$V$_{0.09}$Ti$_{0.06}$Mn$_{0.04}$)$_{3.04}$S$_4$. These sections could not be understood completely by non-specialist readers in Solid State Physics due to the technical terms used, however they can be considered as references for the next sections.

Section 4 focuses on the open issues that make these minerals interesting to investigate.

Section 5 highlights the impact of the results to be obtained when the open issues are addressed in future experiments. In this section the term Technosignature will be described as a recent emerging field in Planetary Science.

Section 6 summarizes the open issues and the recommendations for how to proceed next in future experiments to address them.

## 2. SELECTED TRANSITION METAL SULFIDES

### 2.1. Synthetic $Cr_3S_4$, $(Fe, Ti)_3S_4$ and $(Fe, V)_3S_4$ systems

#### 2.1.1. Crystallographic structure

The $(Fe, Ti)_3S_4$ and $(Fe, V)_3S_4$ systems are isostructural with the monoclinic $Cr_3S_4$ type structure as shown in Fig. 1. This structure is metal-deficient NiAs-like, in which there are two crystallographic sites for cations, one site in a metal-deficient layer ($M_I$) with ordered metal vacancies and the second site in a metal full layer ($M_{II}$). These layers are intercalated between hexagonal close-packed sulfur layers. The lattice parameters of $Cr_3S_4$ are a = 5.964 Å, b = 3.428 Å, c = 11.272 Å and $\beta$ = 91.50 ° [1].

In the $(Fe, Ti)_3S_4$ system, two synthetic phases have been reported so far; $FeTi_2S_4$ and $Fe_2TiS_4$. It was evidenced that in $FeTi_2S_4$, Fe atoms are located in $M_I$ layer and Ti atoms in the $M_{II}$, and that in $Fe_2TiS_4$, the $M_I$ layer is occupied by Fe atoms while the $M_{II}$ layer is occupied by Fe and Ti atoms with equal proportions ([2] and references therein). Regarding $(Fe, V)_3S_4$ system, similar results on site distributions of Fe and V atoms are reported, that is, in $FeV_2S_4$, Fe atoms are located in the $M_I$ layer and V atoms in $M_{II}$, and in $Fe_2VS_4$, the $M_I$ layer is occupied by Fe atoms while the $M_{II}$ layer is occupied by Fe and V atoms with equal proportions ([3] and references therein).

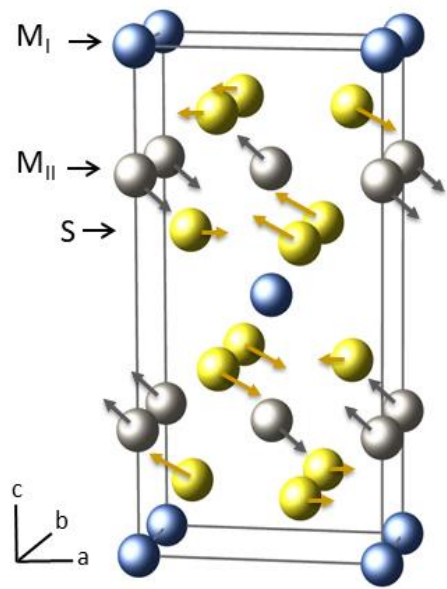

Fig. 1: The ideal structure of $Cr_3S_4$ (Space group I2/m, No. 12) after Jellinek [1], blue spheres: metals in the metal deficient layer ($M_I$), grey spheres: metals in the metal full layer ($M_{II}$) and yellow spheres: sulfur atoms. The arrows indicate directions and magnitudes (five times enlarged) of the observed deviations from the ideal atomic positions, see text.

The (Fe, Ti)$_3$S$_4$ and (Fe, V)$_3$S$_4$ systems have layered structures with different types, depending on the atomic displacements in the Cr$_3$S$_4$ base structure (see Fig. 1). The layered structures are shown in Fig. 2. There is an interesting fact that the atomic displacements are not dependent neither on the nature of metals nor their concentration; that is:

Case (Fe, Ti)$_3$S$_4$ system [2]: the phase FeTi$_2$S$_4$ has almost an "ideal" Cr$_3$S$_4$ type structure with no atomic displacements (type 1, Fig. 2) and the lattice is formed by alternating layers M$_I$ and M$_{II}$ in the [001] plane or basal plane. The atomic separation in M$_{II}$ layer, d$_{22}$, favors the metallic conductivity. In the phase Fe$_2$TiS$_4$ the metals in M$_{II}$ layer have noticeable displacements in the unit cell (type 2, Fig. 2) and the lattice is formed by staircase-like sheets located in the [101] plane, where each sheet is composed by two metallic chains d$_{12}$ and d$_{22}$, in which metallic conductivity is favored. The separation between sheets, d$_{ss}$, is greater than d$_{12}$ and d$_{22}$ and does not favor the metallic conductivity between sheets.

Case (Fe, V)$_3$S$_4$ system [3]: both phases FeV$_2$S$_4$ and Fe$_2$VS$_4$ have staircase-like sheets due to atomic displacements in the unit cell (type 2, Fig. 2), similar to Fe$_2$TiS$_4$ phase.

The fact that the type of layered structure is not correlated to the systems (Fe, Ti)$_3$S$_4$ and (Fe, V)$_3$S$_4$ will be used for the interpretation of microscopic magnetism characterized by $^{57}$Fe Mössbauer spectroscopy below.

### 2.1.2. $^{57}$Fe Mössbauer study

The (Fe, Ti)$_3$S$_4$ and (Fe, V)$_3$S$_4$ systems have been investigated by $^{57}$Fe Mössbauer spectroscopy at variable temperature, in Fig. 3 are shown the spectra of FeTi$_2$S$_4$ and Fe$_2$TiS$_4$ phases at 300 K and 77 K [2]. We can note that, at room temperature (300 K), the spectrum of FeTi$_2$S$_4$ is a doublet that is related to one site for Fe in the M$_I$ layer, and the spectrum of Fe$_2$TiS$_4$ is composed of two doublets that are related to different crystallographic sites; one site in the M$_I$ layer and another in the M$_{II}$ layer. The FeTi$_2$S$_4$ phase undergoes a transition from paramagnetic to magnetic ordering at transition temperature $T_c$ = 145 K, giving rise to an _unusually low_ hyperfine magnetic field (HF) = 2.5 T (at 77 K) if compared with the values of iron magnetic moments (determined from magnetic susceptibility measurements) μ/Fe = 3.40 – 3.60 μ$_B$ and with the associated hyperfine constant in metals of 15 T/μ$_B$. This behavior is explained on the basis of blocking of Fe localized magnetic moments located in the M$_I$ layer by the Spin density wave (SDW) originated from 3d Ti atoms located in the M$_{II}$ layer. In the Fe$_2$TiS$_4$ phase a paramagnetic-SDW transition arises at $T_c$ = 290 K; the SDW spreads in both Fe and Ti atoms through the [101] crystallographic plane, i.e. the staircase-like sheets (type 2 layered structure, see Fig. 2) and undergoes a first order transition from Incommensurate SDW (ISWD) to Commensurate SDW (CSDW) at $T_{IC}$ = 255 K.

In the case of (Fe, V)$_3$S$_4$ system, ISDW has been found for both phases FeV$_2$S$_4$ and Fe$_{1.8}$V$_{1.2}$S$_4$ [3] (see Fig. 4), and the ISDW spreads in both Fe and V atoms through the [101] crystallographic plane, i.e. the staircase-like sheets (type 2 layered structure, see Fig. 2).

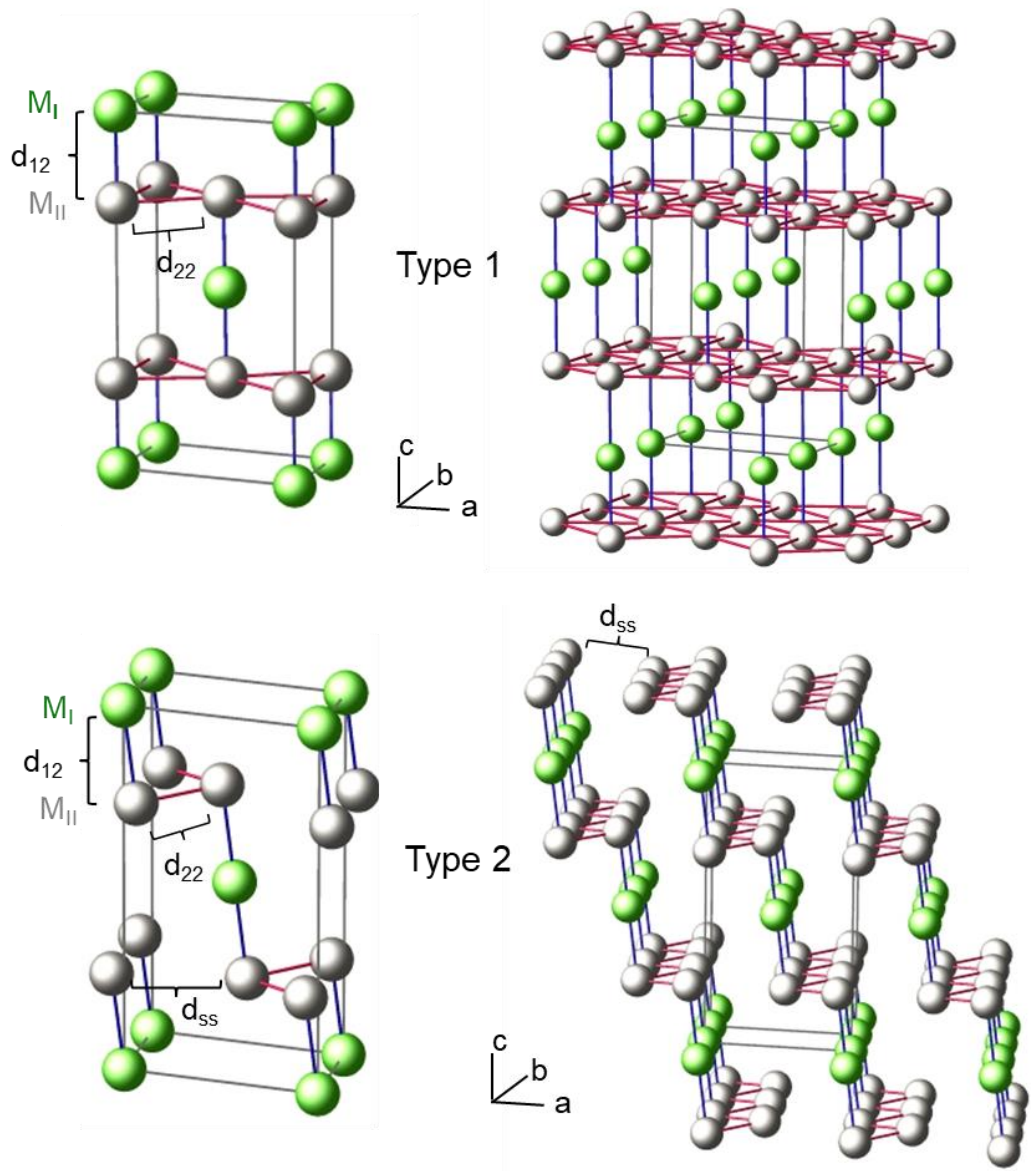

Fig. 2: Layered structures as determined in (Fe, Ti)$_3$S$_4$ and (Fe, V)$_3$S$_4$ systems [2, 3]. The so called "type 1" layered structure (up) with no atomic displacements in the unit cell (left) and the lattice is formed by layers in the basal plane (right). The "type 2" layered structure (down), with atomic displacements in the unit cell and the lattice is formed by staircase-like sheets. Sulfur atoms are omitted for the sake of simplicity, see text.

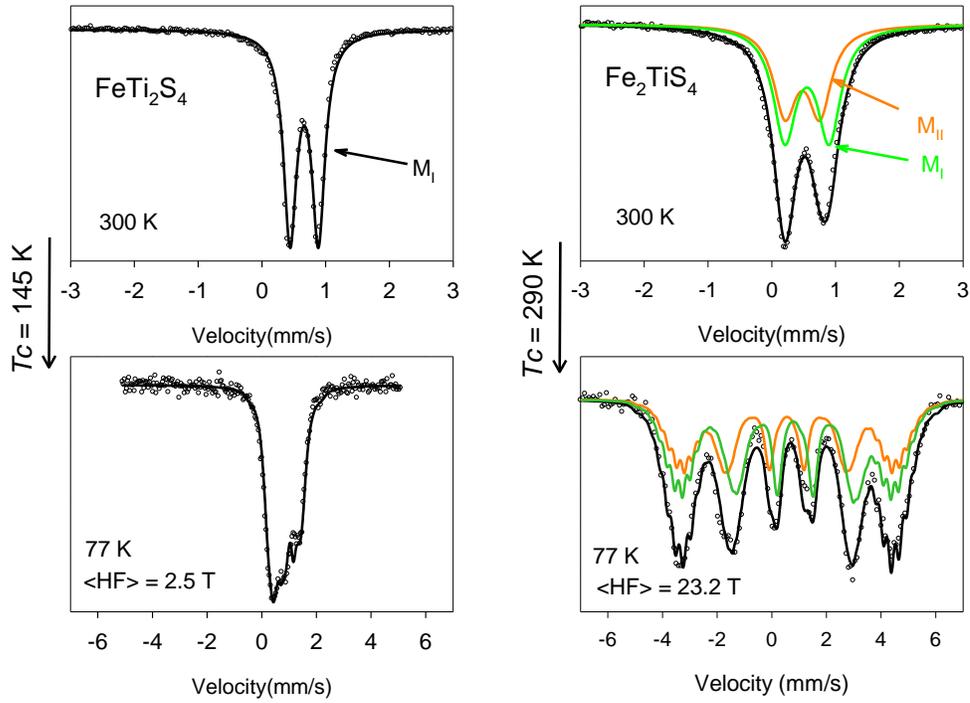

Fig. 3: Mössbauer spectra of $FeTi_2S_4$ and $Fe_2TiS_4$ at 300 K and 77 K, where $Tc$ and <HF> stand for transition temperature and average hyperfine magnetic field [2].

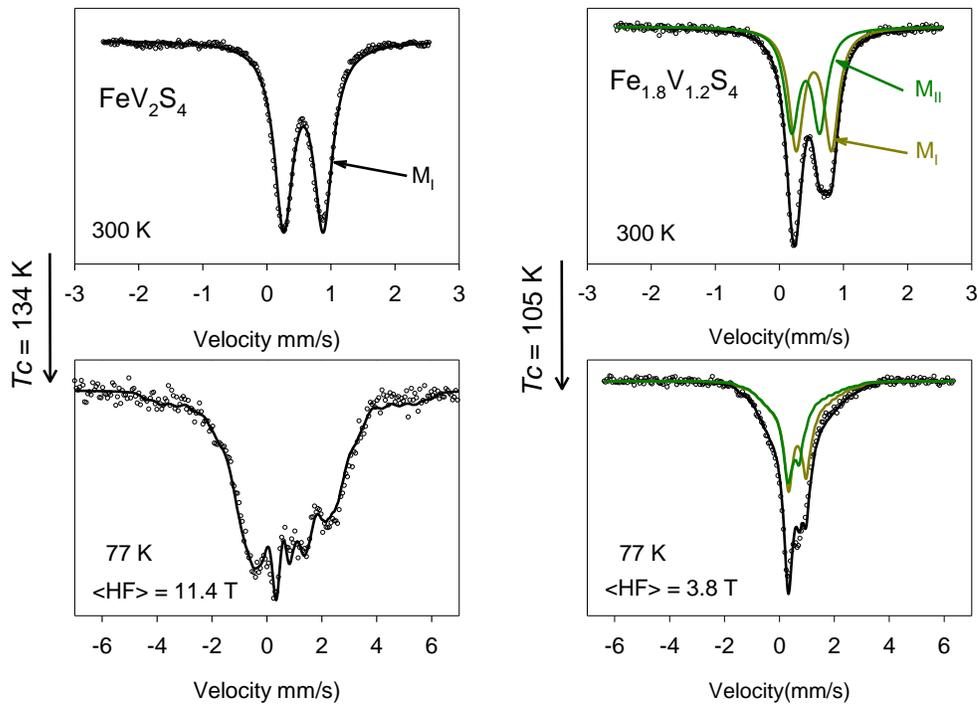

Fig. 4: Mössbauer spectra of $FeV_2S_4$ and $Fe_{1.8}V_{1.2}S_4$ at 300 K and 77 K. Spectra at 77 K are due to ISDW transition [3].

Finally, it is interesting to note that the magnetic behavior with temperature in both systems is "opposed", i.e. *Tc* increases with iron concentration in $(Fe, Ti)_3S_4$ system but deceases in $(Fe, V)_3S_4$ system, and in a similar trend, HF increases at low temperature with increasing iron concentration in $(Fe, Ti)_3S_4$ but deceases in $(Fe, V)_3S_4$ (see Figs. 3 and 4**).**

## 3. THE METEORITIC MINERALS HEIDEITE AND BREZINAITE

### 3.1. Occurrence

The synthesized sulfides mentioned above have no natural occurrence on Earth and were detected as minerals in meteorites years later after the first synthesis; $FeTi_2S_4$, was first synthesized in 1968 [4] and then detected in 1974 as the mineral Heideite in the Bustee meteorite [5] and in 1995 in the Kaidun meteorite [6]. The base structure $Cr_3S_4$ was first synthesized in 1957 [1] and then detected in 1969 as the mineral Brezinaite, with occurrence in the Tucson meteorite [7]. These minerals were structurally indexed as a monoclinic $Cr_3S_4$ type structure using X-Ray Diffraction.

### 3.2. Minor and traces of (Fe, V, Ti, Mn) elements in Brezinaite and Cr in Heideite

The mineral Brezinaite was reported with minor proportion of Fe atoms and traces of V, Ti and Mn atoms using electron microscopy, leading to the composition formula $(Cr_{2.65}Fe_{0.20}V_{0.09}Ti_{0.06}Mn_{0.04})_{3.04}S_4$ [7]. Regarding the mineral Heideite, minor proportions of Cr atoms were detected by electron microscopy and the reported composition formulae are compiled in a general formula $(Fe, Cr)_{1+x}(Ti, Fe)_2S_4$, specifically; $(Fe_{0.99}Cr_{0.16})_{1.15}(Ti_{1.70}Fe_{0.30})_2S_4$ in the Bustee meteorite [5] and $(Fe_{0.88}Cr_{0.41})_{1.29}(Ti_{1.85}Fe_{0.15})_2S_4$ in the Kaidun meteorite [8].

## 4. WHY ARE THESE MINERALS SO INTERESTING?

**Because of the following issues:**

**4.1.** No Mössbauer studies have been carried out on these minerals, among ~ 220 meteorites, analyzed and reported so far in Mössbauer Mineral Handbook [9] and up to date.

**4.2.** These minerals are structurally sensitive to the method of synthesis, and so is their magnetic behavior, especially in the presence of Cr atoms in Heideite and Fe atoms in Brezinaite with V, Ti and Mn traces, that is:

**4.2.1.** The synthetic "pure" systems $(Fe, Ti)_3S_4$ and $(Fe, V)_3S_4$ have the monoclinic $Cr_3S_4$ type structure as stated above (space group I 12/m1, No 12) and all these three sulfides can be synthesized by annealing procedure (heating and slow cooling) [1,4,10-12].

- **4.2.2.** On the other side, the Cr rich system (Fe, Cr)$_3$S$_4$ from which the known reported phase is FeCr$_2$S$_4$ (which is the meteoritic mineral Daubréelite), can be synthesized by a similar annealing procedure but crystallize in cubic spinel structure (space group F d -3 m S, No 227) [13-18]. This phase undergoes a structural transformation to monoclinic type Cr$_3$S$_4$ structure if treated with temperature and pressure, then followed by quenching at room temperature; Bouchard [19] reported the spinel – monoclinic transformation at 1273 K and 65 kb for one hour, Tressler [20] reported the same transformation at 800 K and 55 kb for one week.

- **4.2.3.** So, there is no phase diagram well established for the synthesis of the monoclinic systems leading to the composition formula of Heideite (Fe, Cr)$_{1+x}$ (Ti, Fe)$_2$S$_4$ nor one of Brezinaite (Cr$_{2.65}$Fe$_{0.20}$V$_{0.09}$Ti$_{0.06}$Mn$_{0.04}$)$_{3.04}$S$_4$.

- **4.2.4.** Concerning the magnetic behavior, the presence of minor proportion of Cr atoms in Heideite is quite interesting; it is known that the metallic chromium and some chromium alloys have itinerant antiferromagnetic behavior in the form of SDW [21,22], and the transition temperature to SDW in metallic chromium is about ~313 K. Another particular case is the presence of Mn trace in Brezinaite; the chromium – manganese alloy Cr$_{0.96}$Mn$_{0.04}$ exhibits a transition from CSDW to ISDW with decreasing temperature from 236 K [23]. So, the presence of minor and trace elements of Cr and Mn in Heideite and Brezinaite could affect considerably the commensurability of SDW and the magnetic transition temperature.

## 5. SIGNIFICANCE IN SOLID STATE PHYSICS AND PLANETARY SCIENCE

### 5.1. Solid State Physics: Iron-Based Superconductors?

The structural, magnetic and compositional properties of these meteoritic minerals make them as *promising candidates for* iron-based superconductors because of the following clues:

### 5.1.1. Structural property: Layered structures.

The lattice configuration of Heideite and Brezinaite should be one of layered structures type 1 or type 2 (see sec. 3.1 and Fig. 2). Similarly, iron-based superconductors have layered structures, where superconducting behavior takes place in layers containing irons, such as the case of F-doped LaFeAsO [24] and Co-doped $BaFe_2As_2$ [25] among others. A comprehensive review of lattice configuration and layer types of iron-based superconductors is published elsewhere [26]. In Fig. 5 are shown the structures of the parent (undoped) compounds LaFeAsO and $BaFe_2As_2$.

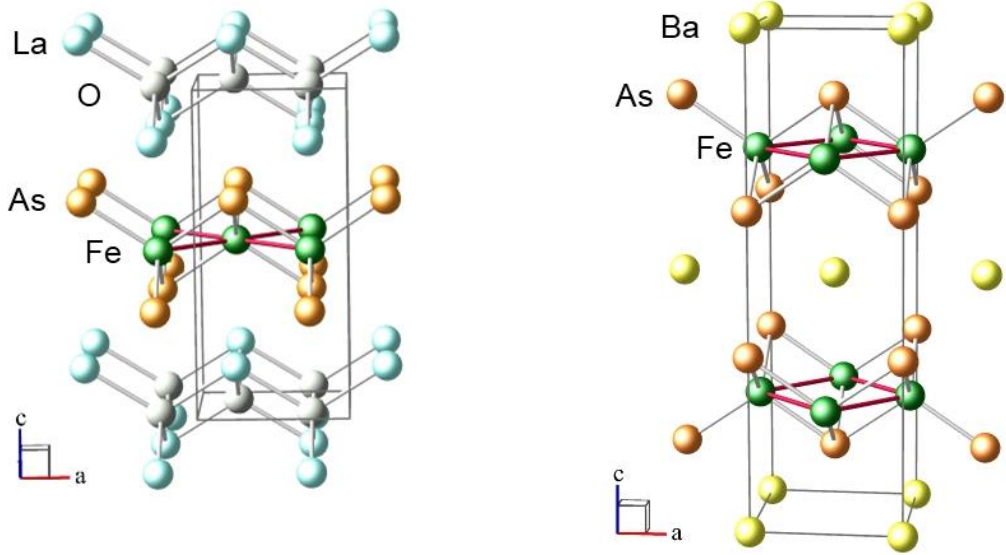

Fig. 5: The structures of LaFeAsO (left) [27] and $BaFe_2As_2$ (right) [28], shown with relative sizes of the unit cell which is tetragonal.

**5.1.2. Magnetic property: transition to spin density wave.**

Parent compounds of Iron-based superconductors undergo a transition to SDW with variable degree of commensurability; in the case of LaFeAsO the transition temperature is $Tc$ = 150 K [29] and in the case of BaFe$_2$As$_2$, $Tc$ = 140 K [30] (see Fig. 6). It is interesting to note the comparison with the synthetic sulfide systems (Fe, Ti)$_3$S$_4$ and (Fe, V)$_3$S$_4$ (see Figs. 3 and 4); the spectrum at 77 K of BaFe$_2$As$_2$ is due to CSDW which is similar to that of Fe$_2$TiS$_4$ phase (at 77 K) and the spectra at (130 -140) K exhibit ISDW behavior which is similar in both FeV$_2$S$_4$ and Fe$_{1.8}$V$_{1.2}$S$_4$ phases.

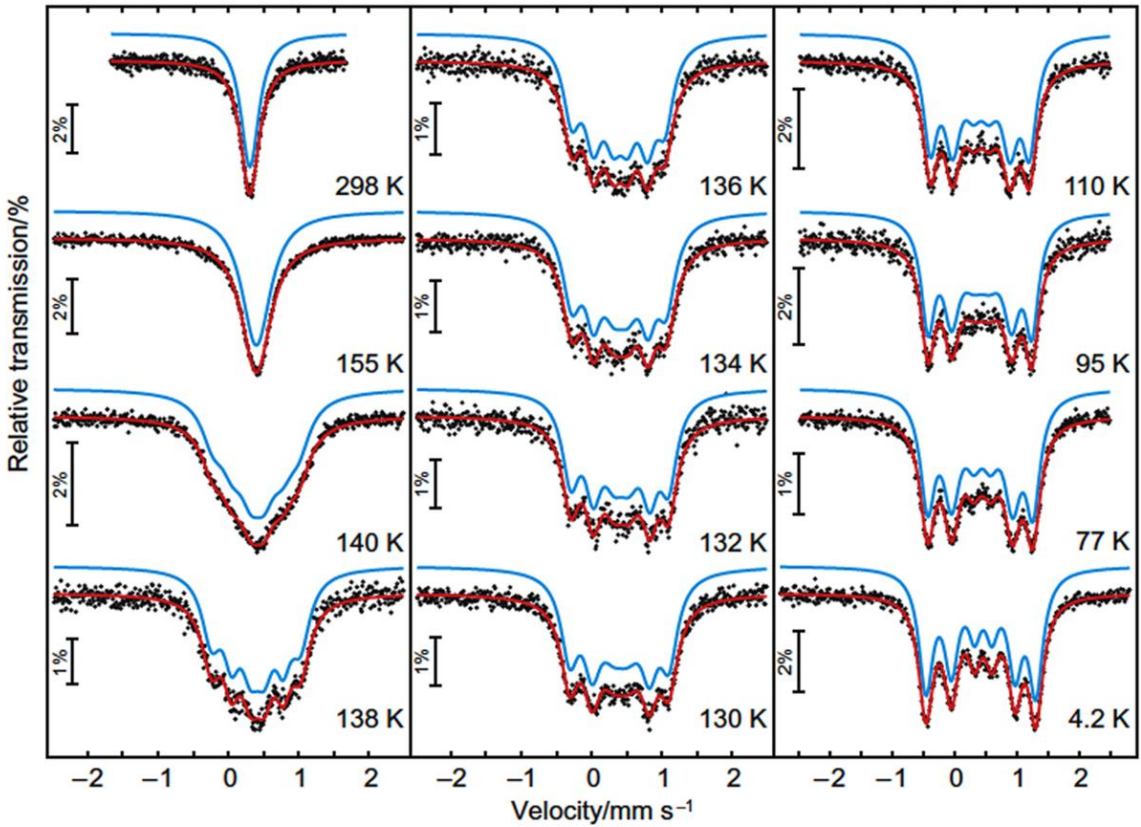

Fig. 6: Mössbauer spectra of BaFe$_2$As$_2$, blue lines represent transmission integral fits. Figure reprinted from Rotter M., et al [30]. *Copyright IOP Publishing Ltd and Deutsche Physikalische Gesellschaft. Reproduced by permission of IOP Publishing.* [CC BY-NC-SA](CC BY-NC-SA).

### 5.1.3. Compositional property: minor and trace elements as dopants.

When parent compounds of Iron-based superconductors are doped with certain minor or trace element, the SDW is eventually suppressed or loses commensurability and superconductivity emerges; such as the case of F-doped LaFeAsO (called hole doping) where SDW is completely suppressed [31] and the case of Co-doped $BaFe_2AS_2$ (called electron doping) where SDW undergoes a CSDW – ISDW transition [25], (see Fig. 7).

Regarding the presence of minor and trace elements like Cr in Heideite and Mn in Brezinaite (see sec. 4.2.4) it is naturally interesting to investigate the role of these elements as possible dopants and hence leading to superconducting behavior.

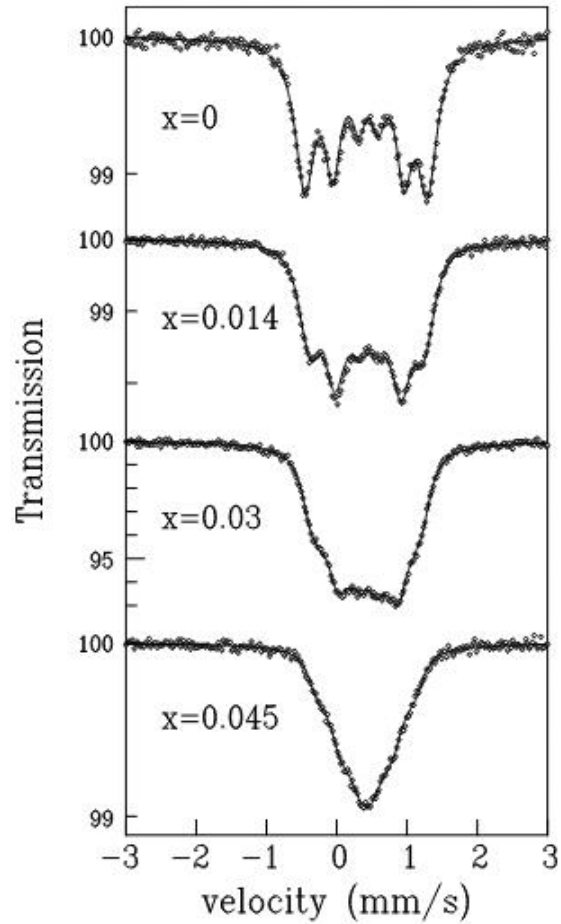

Fig. 7: Mössbauer spectra of $Ba(Fe_{1-x}Co_x)_2As_2$ at 4.2 K for x = 0, 0.014, 0.03 and 0.045. Except for x = 0, all spectra are due to ISDW. Figure reprinted from Bonville P., et al [25]. *Copyright EPLA. Reproduced by permission of Europhysics Letters.*

## 5.2. Planetary Science: Technosignatures?

The genesis of these meteoritic minerals could require controlled and sophisticated process not easily found in nature (see secs. 4.2.1 - 4.2.3). So, we can go a step ahead and postulate the hypothesis of non-natural or artificial origin of these meteoritic minerals. More precisely; they are possible Technosignatures on Earth. Of course, this hypothesis needs to be addressed by experimental procedures, from which the results should be analyzed in the framework of Planetary Science. A brief review about the emerging field of Technosignatures is given in the following.

### 5.2.1. What are Technosignatures?

Technosignatures are defined as physical manifestations of technology from extraterrestrial intelligences; this definition was first coined by Jill Tarter at SETI (search for extraterrestrial intelligence) Institute [32]. In September 2018, NASA hosted the "NASA Technosignatures Workshop"[1] to discuss about the current state of art for technosignature searches. Following this event, in December 2018; US Congress authorized funding[2] for each of fiscal years 2018 and 2019 to search for technosignatures. And more recently, many researchers suggested strategies for technosignature searches [33-39]. This emerging field is so wide because of the variety of technosignature types to search for [38]. Hence, an interdisciplinary research team could be decisive to achieve promising results [33-35,37]. In this context, this review could make a contribution to this interesting field from the Solid State Physics approach.

The technosignature type for the meteoritic minerals Heideite and Brezinaite should belong to the category of Solar system artifacts [38]. Each type of technosignature requires a proper search strategy. There is a timely framework proposed by Sheikh [36] to compare the viabilities or merits of different search strategies, based on nine variables called "nine axes of merit".

We present in the following the nine axes of merit for the meteoritic minerals, based on similar criteria used by Sheikh [36] in the evaluation for Solar system artifacts, with slight qualitative modification, given that the meteoritic minerals contained in the meteorites Bustee, Kaidun and Tucson are present on Earth, and probably belong to the subcategory of "derelict technology" as defined by Wright [39].

### 5.2.2. Nine Axes of Merit

The nine axes of merit used to justify the search for meteoritic minerals as technosignatures are evaluated below (see **Fig. 8**):

#### 5.2.2.1. Observing capability

---

[1] https://doi.org/10.48550/arXiv.1812.08681

[2] https://www.congress.gov/115/crpt/hrpt1102/CRPT-115hrpt1102.pdf

There are unanswered questions – a puzzle - about these minerals that deserves experimental attention in the attempt to be solved. This exciting opportunity for new experimental research in this area include the design and implementation of the appropriate (and challenging) synthesis method leading to analog samples of these meteoritic minerals, their characterization with $^{57}$Fe Mössbauer Spectroscopy (see sec. 4.1), X-Ray Diffraction and Resistivity. These experiments belong to our technological capability for more than two decades, and can be carried out at any time, when funding is available. (Score = 100).

### 5.2.2.2. Cost

The cost is relatively low, it basically covers the synthesis of analogue minerals, the instruments time when used, i.e., Mössbauer Spectrometer, X-Ray Diffractometer and Resistivity measurements. (Score = 90).

### 5.2.2.3. Ancillary Benefits

The possible and prominent ancillary benefit is technological; the development of new class of iron-based superconductors (see sec. 5.1). Iron-based superconductors are extremely demanded since they seem to have potentially useful technological applications due to their relative high Tc superconducting state. For example; the actual superconducting devices such as superconducting magnets used in Magnetic Resonance Imaging (MRI), have Tc in order of ~ 10 K, and they are cooled with liquid helium to maintain temperatures near 4 K. Helium is a very rare and expensive element. On the other side, liquid nitrogen (T = 77 K) is not expensive because the nitrogen element is the main component of the atmosphere (78 %) and superconductors with Tc > 77 K may be cooled with liquid nitrogen. (Score = 80).

### 5.2.2.4. Detectability

The meteoritic minerals are contained in the meteorites Bustee, Kaidun and Tucson. If we suppose that these meteorites as examples of solar system artifacts [38], then the detectability is almost maximum, because they are on Earth. However, this evaluation is strongly dependent on the research to carry out in the first merit of Observing capability. There are interesting and intriguing clues about the genesis of these minerals that would be verified by the experimental results to be obtained. (Score = 70).

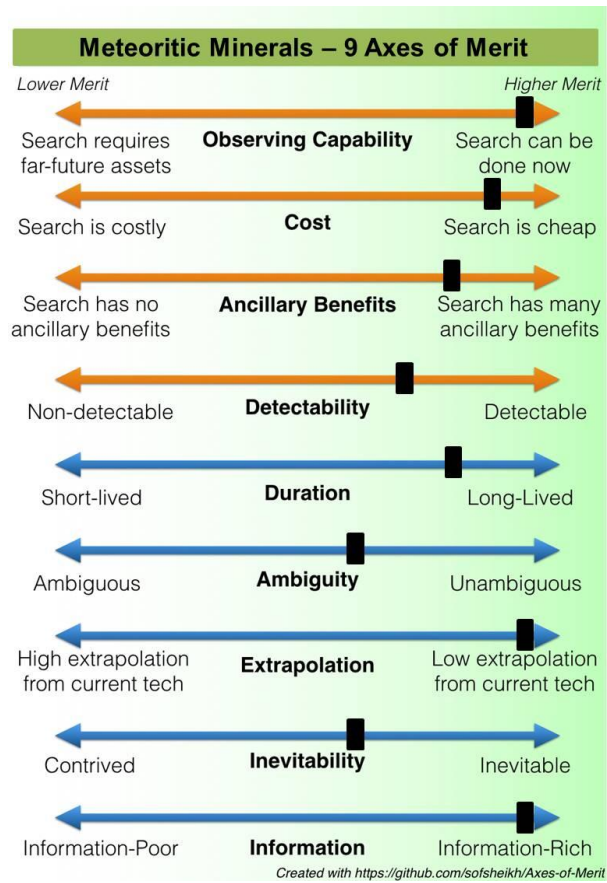

**Fig. 8**: Nine Axes of Merit for Meteoritic Minerals Hedeite and Brezinaite. The plot was generated using open software created by Sheikh [36] used to compare searches for different kinds of technosignatures.

### 5.2.2.5. Duration

In the same reasoning as previous axis of Detectability, the meteoritic minerals are on Earth since decades ago, so, if the experimental results prove that they are technosignatures, the duration should reach the maximum of the axis, the experiment is decisive in this case. (Score = 80).

### 5.2.2.6. Ambiguity

The term Ambiguity depends on how the synthesis method leading to analog samples of these meteoritic minerals could be explained in the context of Planetary Science for the formation of these minerals; it could be unambiguous if there is no satisfactory explanation about the "natural" synthesis of these minerals. (Score = 60).

### 5.2.2.7. Extrapolation

The two interesting and intriguing facts about these meteoritic minerals are their genesis and possible superconducting behavior. These facts are based in our current scientific and technological background in Solid state physics. So the Extrapolation value favors the study of these minerals as possible technosignatures because of low or almost zero extrapolation from our current technology. (Score = 100).

### 5.2.2.8. Inevitability

It is suggested that the Inevitability score for Solar system artifacts should lie near the minimum (contrived) [36], based on the argument that the there is no physically motivated reason for the construction of interstellar artifacts to be a common phenomenon. However, this argument could not limit the probable existence of these artifacts on Earth, coined as "Alien Artifacts" [40], "Footprints of Alien Technology" [41] and "Non-Terrestrial Artifacts" [42]. (Score = 60).

### 5.2.2.9. Information

The information to be obtained is summarized in the following three questions that deserve experimental attention to be answered:

**Question 1:** What if we cannot obtain analog samples of theses minerals, as they are, with minor and trace metals? What is the implication about their genesis?

**Question 2:** Whatever the analog samples to be obtained, are they superconductors? If yes, what is the superconducting transition temperature?

**Question 3:** Given the plausible answers to the above questions, we feel it is important to be open-minded and even provocative – to consider the following question; Are these meteoritic minerals samples of Extraterrestrial Technosignatures? (Score = 100).

## 6. CONCLUSIONS AND OUTLOOK

The meteoritic minerals Heideite $(Fe, Cr)_{1+x} (Ti, Fe)_2 S_4$ and Brezineite $(Cr_{2.65}Fe_{0.20}V_{0.09}Ti_{0.06}Mn_{0.04})_{3.04}S_4$ have interesting and intriguing key issues that emerged following the experimental results reported so far for synthetic sulfide systems $(Fe, V)_3 S_4$ and $(Fe, Ti)_3 S_4$, which are isostructural with these minerals. The open issues are related with the composition of these minerals and deserve future experiments in the attempt to be solved; i.e., the Solid State Physics issue about their magnetic and transport properties and the Planetary Science issue about their genesis.

The interest to carry out future experiments relies in the significance of the results to be obtained; in Solid State Physics these minerals are promising candidates for iron based superconductors, and in Planetary Science the design and implementation of an appropriate (and challenging) synthesis method, leading to analog samples of these meteoritic minerals will shed light on their genesis. So, it is timely and provocative the following question: Are these minerals samples of Extraterrestrial Technosignatures?

Whatever the results to be obtained from the experiments, they constitute a contribution to our knowledge in Solid State Physics and Planetary Science; this is the essence of the Scientific Method. The experiments could be conducted by multidisciplinary team of Solid State Physicists specialized in synthesis and characterization techniques such as $^{57}$Fe Mössbauer Spectroscopy, Powder X-Ray Diffraction and Resistivity, and Planetary Scientists specialized in Meteoritics. Also, and subject to their availability, it is desirable to obtain samples of the meteoritic minerals to characterize them by $^{57}$Fe Mössbauer Spectroscopy.

There are more specific open questions in Solid State Physics about these minerals, not discussed in this manuscript, for the sake of simplicity. However these minor questions will be addressed by the future experiments to carry out.

## ACKNOWLEDGMENTS

The author is grateful to the colleague Jacob Haqq-Misra (Blue Marble Space Institute of Science, Seattle, WA, USA) and Jason Wright (The Pennsylvania State University, PA, USA) for their fruitful discussions and comments on a previous version of the manuscript. The author also thanks the following colleagues for allowing reprinting figures; Dirk Johrendt (Ludwig-Maximilians-Universität München, Germany), (Fig. 6) and Pierre Bonville (Commissariat a l'Energie Atomique (CEA), Saclay, France), (Fig. 7).